# NEW GENERATION ELECTRON-POSITRON FACTORIES

Mikhail Zobov on behalf of DAΦNE, SuperB and SuperC-Tau Accelerator Teams

*INFN, Laboratori Nazionali di Frascati, via Enrico Fermi 40, 00044 Frascati (RM), Italy*

**Abstract**

In 2010 we celebrate 50 years since commissioning of the first particle storage ring ADA in Frascati (Italy) that also became the first electron-positron collider in 1964. After that date the particle colliders have increased their intensity, luminosity and energy by several orders of magnitude. Namely, because of the high stored beam currents and high rate of useful physics events (luminosity) the modern electron-positron colliders are called "factories". However, the fundamental physics has required luminosities by 1-2 orders of magnitudes higher with respect to those presently achieved. This task can be accomplished by designing a new generation of factories exploiting the potential of a new collision scheme based on the Crab Waist (CW) collision concept recently proposed and successfully tested at Frascati. In this paper we discuss the performance and limitations of the present generation electron-positron factories and give a brief overview of new ideas and collision schemes proposed for further collider luminosity increase. In more detail we describe the CW collision concept and the results of the crab waist collision tests in DAΦNE, the Italian Φ-factory. Finally, we briefly describe most advanced projects of the next generation factories based on the CW concept: SuperB in Italy, SuperKEKB in Japan and SuperC-Tau in Russia.

В 2010 году исполнилось 50 лет со дня запуска во Фраскати (Италия) первого накопителя ADA, который в 1964 году стал также первым в мире электрон-позитронным коллайдером. За прошедшее время интенсивность пучков, светимость и энергия коллайдеров увеличились на несколько порядков. Современные электрон-позитронные коллайдеры часто называют "фабриками" из-за большого тока пучков и очень высокой светимости. Однако, для дальнейшего продвижения в физике высоких энергий, требуется увеличить светимость еще на 1-2 порядка. Эта задача может быть решена созданием следующего поколения фабрик, использующих потенциал новой схемы столкновения пучков, которая получила название Crab Waist (CW). Концепция CW была недавно (в 2006 году) предложена во Фраскати, и там же успешно реализована. В этой статье мы обсуждаем производительность современных электрон-позитронных фабрик, чем она ограничивается, и делаем краткий обзор новых идей и схем столкновения пучков, которые были предложены для дальнейшего увеличения светимости. Более детально рассмотрена концепция CW и результаты ее практического применения на DAΦNE, итальянской Φ-фабрике. В заключение мы кратко рассмотрим наиболее продвинутые проекты фабрик нового поколения, основанных на концепции Crab Waist: SuperB в Италии, SuperKEKB в Японии и SuperC-Tau в России.

# Introduction

Since their invention in the beginning of the past century particle accelerators are widely used in many fields of fundamental physics from elementary particles, astrophysics and cosmology to solid state, nuclear and atomic physics. They are essential instruments for medicine, biology, chemistry and also used nowadays for food preservation and sterilization as well as in the elements analysis, archeology and many other applications in our everyday life.

The accelerators for high energy physics exploiting colliding particle beams, "colliders", are indispensable tools for deep studies of the matter (and anti-matter) microstructure aimed at understanding the origin and development of the Universe. The colliders have a remarkable kinematic advantage with respect to fixed target machines. Describing this advantage G.K.O'Neill, one of the collider pioneers, writes in 1956 [1]: "..As accelerators of higher and higher energy are built, their usefulness is limited by the fact that the energy available for creating new particles is that measured in the center-of-mass system of the target nucleon and the bombarding particle. In the relativistic limit, this energy rises only as the square root of the accelerator energy. However, if two particles of equal energy traveling in opposite directions could be made to collide, the available energy would be twice the whole energy of one particle..." We can also add that the colliders are "cleaner" machines with respect to the fixed target ones since the colliding beams do not interact with the target materials. Besides, it is much easier to organize collisions of beams composed of matter-antimatter particles, like in electron-positron and proton-antiproton colliders.

It is believed that the idea of colliding beams belongs to Rolf Wideroe who obtained a patent on this technique in 1953 [2]. But, as Wideroe says in his memories, the idea came him much earlier, in 1943 [3]. Nevertheless, the first serious design proposals for a collider appeared only in 1956 [1, 4]. A group at the Midwestern Universities Research Association (MURA) led by D.W.Kerst proposed building for this purpose two tangent fixed-field accelerators having a common straight section for beam collisions [4]. In the same year G.K.O'Neill suggested using a single accelerator to inject particles into a pair of tangent storage rings [1]. The benefit of storage rings consists in that the storage ring can accumulate and bring into collision beams with much higher intensities.

Soon after that many groups in several laboratories started working on colliding beams and, almost at the same time, the first colliders came into operation in USA, Soviet Union and Italy in the early 1960s. Construction of the Princeton-Stanford electron-electron collider started in 1959 [5] while the first electron-electron collisions were obtained in 1965 and the first interesting results were published in 1966 [6]. The first Soviet e⁻e⁻ storage ring, VEP1, was constructed in Moscow and moved to Novosibirsk in 1962 [7, 8]. In 1965 VEP1 started giving first experimental results [9, 10]. Before the success of the electron-electron colliders, an Italian group at Laboratori Nazionali di Frascati led by Bruno Touschek designed and built the first storage ring ADA (Anello di Accumulazione), proved the possibility of storing an accelerated beam for hours [11, 12] and accomplished first electron-positron collisions in 1964 [13]. ADA was the first electron-positron collider.

The interest to electron-positron colliders was growing. Soon after ADA other small low energy colliders became operative: VEPP2 in Novosibirsk (Soviet Union) [14] and ACO in Orsay (France) [15]. Despite their small energy and sizes and relatively low luminosity, the first colliders gave significant contributions in particle physics and helped to discover and explain many accelerator physics phenomena. The first step towards higher energy colliders was made designing [16] and commissioning the electron-positron collider ADONE at Frascati Laboratories in 1969 [17].

Since then colliders became the leading tool in particle physics research and their scale grew rapidly both in energy and luminosity. Besides, the variety of colliding particles kinds has been expanding. Here we list just a few examples: electron-proton collisions in HERA at DESY [18], proton-antiproton collisions at Tevatron at Fermilab [19], proton-proton collisions in ISR and LHC at CERN [20], ion-ion collisions in RHIC at Brookhaven [21] and LHC etc. However, in this article we will focus only on the electron-positron colliders. Figure 1 summarizes the peak luminosities and energies of the past, present and future e+e- colliders.

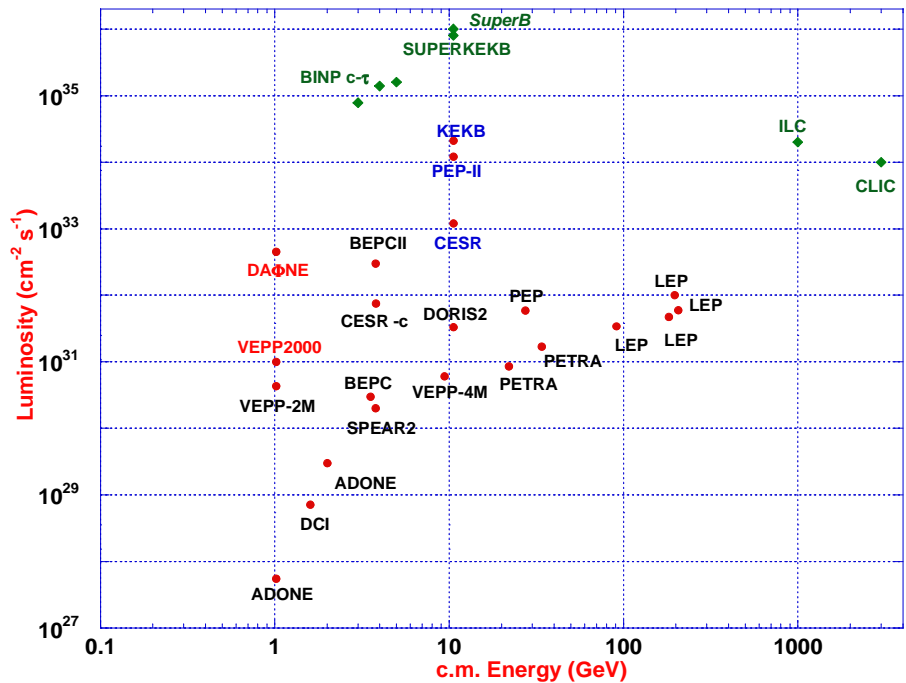

Figure 1. Peak luminosity and energy of the past, present and future (diamonds) electron-positron colliders.

LEP, operated at CERN [22], was the biggest collider (27 km circumference) having the highest energy ever achieved in an e+e- collider (a top energy of 106 GeV per beam). Most probably LEP was the last step towards higher energies in the electron-positron colliders based on storage rings because of the dramatic rise in synchrotron radiation loss with energy increase. Indeed, an enormous RF system using many superconducting cavities was necessary to compensate the huge energy loss in LEP, as high as 3 GeV per revolution turn. In order to proceed with energy increase in electron-positron collisions it is planned to use linear colliders, such as ILC [23] and CLIC [24], for example. Certainly, the linear colliders deserve a separate discussion, but it is out of the scope of our paper.

The luminosity increase is another frontier, another direction to fully exploit the potential of circular electron-positron colliders. The fundamental physics required a substantial step in luminosity increase for precision measurements of rare decays, extremely small cross-sections, CP violation events etc. Narrow energy regions corresponding to the quark resonances are of particular interest for these physics studies. In order to explore the narrow resonances, dedicated meson "factories", the electron-positron factories with very high luminosity, were designed and built in the 1990s. These are: the Φ-factory DAΦNE [25] in Italy, PEPII [26] and KEKB [27] the B-factories in USA (Stanford) and Japan (Tsukuba), respectively. Recently also a Tau-Charm factory, BEPCII, came into operation in Beijing (China) [28]. Here it is worthwhile saying that, in a certain sense, LEP and CESR in Cornell [29] can be considered as "prototypes" for the present generation factories which demonstrated the feasibility of multibunch collisions with trains of bunches using Pretzel scheme to minimize the effect of parasitic interactions [see discussion in 30, for example].

The present generation of e+e- factories has been very successful in reaching their design goals in terms of stored beam currents, peak and integrated luminosities and in developing and testing many new accelerator techniques and innovative accelerator technologies. However, interests of physicists and experimentalists go much beyond the presently achieved luminosity levels. Work aimed at finding new ways for luminosity increase is always in progress. Several new ideas and novel collision schemes have been proposed and some of them tested at the electron-positron factories. At present a recently proposed collision scheme, called "Crab Waist (CW)" [31], is considered most promising for designing a new generation of $e^+e^-$ factories with luminosities by 1-2 orders of magnitude higher with respect to those obtained in the present factory-class machines. The advantages of the CW concept and its successful test at the Italian Φ-factory DAΦNE [32] have already given rise to several new generation factory proposals and upgrade of some existing e+e- factories.

In this review article we discuss the strategy of the present generation electron-positron factories in achieving high luminosity, describe their current performance and intrinsic limitations. Then we give a brief overview of new ideas and novel collision schemes proposed for further collider luminosity increase. In more detail we describe the CW collision concept and the results of the crab waist collision tests at DAΦNE. Finally, we briefly overview the most advanced projects of the next generation factories based on the CW concept: SuperB in Italy [33], SuperKEKB in Japan [34] and SuperC-Tau in Russia [35].

**Present Lepton Factories: Performance, Strategy and Limitations**

Present generation lepton factories have been very successful (see Table 1). Both B-factories, KEKB in Japan and PEPII in USA, have largely exceeded their design goals. The Italian Φ-factory DAΦNE has exceeded the phase I design luminosity and obtained a luminosity increase by a factor 3 after implementation of a novel crab waist collision scheme (discussed below). The recently commissioned Tau-Charm factory in Beijing is well advanced on the way to obtain its design luminosity.

Table 1. Electron-Positron Factory Luminosities

| Factories | Location | Design Luminosity | Achieved Luminosity |
|---|---|---|---|
| KEKB | B-Factory KEK, Japan | $1.0 \times 10^{34}$ | $2.1 \times 10^{34}$ |
| PEP-II | B-Factory SLAC, USA | $3.0 \times 10^{33}$ | $1.2 \times 10^{34}$ |
| DAΦNE, phase I | Φ-Factory Frascati, Italy | $1.0 \times 10^{32}$ | $1.6 \times 10^{32}$ |
| DAΦNE, upgrade | Φ-Factory Frascati, Italy | $5.0 \times 10^{32}$ | $4.5 \times 10^{32}$ |
| BEPCII | Tau-Charm-Factory Beijing, China | $1.0 \times 10^{33}$ | $3.3 \times 10^{32}$ |

As it is seen in Table 2, the high luminosities have been achieved bringing into collision beams with world record intensities. Indeed, PEPII has the largest positron current ever achieved in a storage ring. DAΦNE has the highest electron beam current among operating colliders and synchrotron light sources, while KEKB was capable to accumulate both electron and positron beam currents above 1 A with superconducting cavities.

Table 2. Record beam currents stored in the electron-positron factories.

| Parameters | PEP-II | | KEKB | | DAΦNE | |
|---|---|---|---|---|---|---|
| | LER | HER | LER | HER | e+ | e- |
| Circumference, m | 2200 | 2200 | 3016 | 3016 | 97.69 | 97.69 |
| Energy, GeV | 3.1 | 9.0 | 3.5 | 8.0 | 0.51 | 0.51 |
| Damping time, turns | 8000 | 5000 | 4000 | 4000 | 110000 | 110000 |
| Beam currents, A | 3.21 | 2.07 | 1.70* | 1.25* | 1.40 | 2.45 |

* 2.00 A and 1.40 A were stored in KEKB without crab cavities

In addition to the record stored currents and achieved high luminosities the factories have made many important contributions to the accelerator physics and technology:

- Development of technology for key accelerator components such as RF cavities, both warm and superconducting, innovative vacuum chambers and diagnostics elements;
- Experience in handling multi-ampere currents with powerful feedback systems;
- Top-up and continuous trickle-charge injection with manageable background level;

- Design of interaction regions with permanent and superconducting magnets;
- Operation with crossing angle, crab cavities and crab waist collisions;
- Test and exploitation of techniques for electron cloud suppression;
- Developments and benchmarking of dedicated numerical codes;
- Studies of all beam dynamics aspects: beam coupling impedances, instabilities, beam-beam interactions, and nonlinear dynamics;
- Many others.

A collection of articles summarizing the lepton collider performances can be found in [36].

All the present generation factories relied, at least at the beginning of their operation, on the standard strategy in choosing beam parameters in order to achieve high luminosity. The strategy can be understood by considering the well-known expressions for the luminosity $L$ and beam-beam tune shifts $\xi_{x,y}$ that characterize the strength of harmful nonlinear electromagnetic interaction of the colliding beams. For simplicity we start with the case of head-on collisions of short bunches having equal beam parameters at the interaction point (IP) (see [37], for example):

$$L = N_b f_0 \frac{N^2}{4\pi \sigma_x^* \sigma_y^*} = N_b f_0 \frac{\pi \gamma^2 \xi_x \xi_y \varepsilon_x}{r_e^2 \beta_y^*} \left(1 + \frac{\sigma_y^*}{\sigma_x^*}\right)^2$$

$$\xi_{x,y} = \frac{N r_e}{2\pi \gamma} \frac{\beta_{x,y}^*}{\sigma_{x,y}^* (\sigma_x^* + \sigma_y^*)}$$

(1)

Neglecting beam dynamics aspects, we see from (1) that the luminosity increase in a collider at a given energy requires:

- higher number of particles per bunch $N$
- more colliding bunches, $N_b$
- larger beam emittance, $\varepsilon_x$
- smaller beta functions at the IP, $\beta_y^*$
- beams with equal rms sizes at IP, $\sigma_y^* = \sigma_x^*$
- higher tune shift parameters, $\xi_{x,y}$.

The present factories have obtained their good luminosity performances trying to fulfill almost all the above conditions as much as possible except that:

- it was chosen to collide flat bunches $\sigma_y^* << \sigma_x^*$ since it is rather difficult to provide a good

dynamic aperture for the round beam case with both vertical and horizontal beta functions low at the IP;

- besides, in order to eliminate parasitic collisions in multibunch operation a small horizontal crossing angle $\theta$ was necessary. In the factories a relatively small Piwinski angle $\Phi=(\sigma_z/\sigma_x)tg(\theta/2)<1$ was mandatory to avoid excessive geometric luminosity reduction and to diminish the strength of synchrobetatron resonances arising from beam-beam interaction with the crossing angle.

However, a further substantial luminosity increase based on the standard collision scheme is hardly possible due to several limitations imposed by beam dynamics requirements:

- In order to minimize the luminosity reduction due to the hour-glass effect (the dependence of the vertical beam size on the longitudinal position along the crossing region) the vertical beta function at the IP can not be much smaller than the bunch length;

- A drastic bunch length reduction is impossible without incurring into single bunch instabilities: bunch lengthening and microwave instabilities due to the beam interaction with the surrounding vacuum chamber. Besides, too short bunches tend to produce coherent synchrotron radiation (CSR) affecting beam quality and leading to a dramatic increase of the power losses;

- A multibunch current increase would result in different kinds of coupled bunch beam instabilities, excessive power loss due to interactions with parasitic higher order modes (HOM) and increase of the required wall plug power;

- Higher emittances conflict with stay-clear and dynamic aperture limitations, require again higher currents to exploit the emittance increase for the luminosity enhancement;

- Tune shifts saturate and beam lifetime drops due to a strong nonlinear beam-beam interaction.

Further luminosity increase has required new ideas and nontraditional strategies in beam-beam collisions.

**New Collision Concepts**

In order to overcome the limitations of the standard collision strategy several novel collisions concepts and new collision schemes were proposed. The most known are following:

- Round beam collision preserving an additional integral of motion (see, for example, [38]);

- Crab crossing [39, 40];

- Collision with large Piwinski angle [41] ("superbunch" in hadron colliders [42, 43]);

- Crab waist collision [31, 44, 45];

- Collision with traveling waist [46];

- Longitudinal strong RF focusing [47].

The idea of round beams collision was proposed more than 20 years ago for the Novosibirsk Φ-factory design. It requires equal emittances, equal small fractional tunes, equal beta functions at the IP, no betatron coupling in the arcs. 90° rotation at each passage of the transverse oscillation plane by means of solenoids in the interaction regions (IR) provide conservation of the longitudinal component of the angular moment $M_z = yp_x - xp_y$. Thus the transverse motion becomes one-dimensional. In addition to the obvious advantages coming from (1), the round beam scheme helps to eliminate all betatron coupling resonances that are of crucial importance for tune shift saturation and lifetime degradation. The synchrobetatron resonances are also weakened since the transverse tune shift is almost independent of particle's longitudinal position. The round beam concept was successfully tested at the electron-positron collider VEPP2000 in 2007-2010 at the energy of 510 MeV [48]. Despite the low energy a high single bunch luminosity of $10^{31}$ cm$^{-2}$s$^{-1}$ was achieved together with a maximum tune shift as high as 0.18. Another round beam collisions scheme, "Mobius accelerator", was proposed in [49] and tested at CESR providing a tune shift of 0.09 in agreement with simulations [50].

The crab crossing collision scheme was proposed by R. Palmer in 1988 [39] and further developed in [40]. This idea makes it possible to collide bunches at a large crossing angle without luminosity loss and excitation of synchrobetatron resonances. In the crab crossing scheme both bunches are tilted before collision by half the crossing angle $\theta/2$, providing head-on collision at the IP. The tilt is created by a transverse RF deflector (crab cavity) giving opposite transverse kicks to the bunch head and tail. The RF deflector is placed at a point where the betatron phase in the crossing plane is $-\pi/2$ from the IP. In the classic crab crossing scheme another RF deflector after the collision point is used to restore the tilt. The crab crossing collision, with a single crab cavity per ring, was successfully performed at the KEK B-factory [51]. A world record luminosity of $2.1 \times 10^{34}$ cm$^{-2}$s$^{-1}$ has been obtained in this configuration. However, the achieved luminosity is still lower than that predicted by numerical simulation and work is in progress to find out the reasons of the disagreement.

The idea of colliding with a large Piwinski angle is not a new one as well. In 1995, discussing beam-beam interactions with a large crossing angle, K.Hirata suggested that a large angle might have several merits for future high-luminosity colliders [41]. It has been also proposed for hadron colliders to increase the bunch length and the crossing angle [42, 43] for luminosity optimization. The advantages of a large $\Phi$ can be understood by writing down the formulae for the luminosity and tune shifts with a horizontal crossing angle. Neglecting the hour-glass effect, the expressions can be obtained from (1) simply by substituting the horizontal beam size $\sigma_x^*$ by the effective transverse size $\sigma_x^*(1+\Phi^2)^{1/2}$. Then, for large Piwinski angle, $\Phi \gg 1$, the luminosity and the tune shifts scale as [52]:

$$L \propto \frac{N\xi_y}{\beta_y^*}; \quad \xi_y \propto \frac{N\sqrt{\beta_y/\varepsilon_y}}{\sigma_z \theta}; \quad \xi_x \propto \frac{N}{(\sigma_z \theta)^2} \tag{2}$$

Clearly, in such a case, if it were possible to increase $N$ proportionally to $\sigma_z \theta$, the vertical tune shift $\xi_y$ would remain constant, while the luminosity would grow proportionally to $\sigma_z \theta$. Moreover, the horizontal tune shift would drop proportionally to $1/\sigma_z \theta$.

The idea of using a "travelling" waist (focus) to compensate the luminosity reduction due to the hour-glass effect in circular colliders came from linear colliders [53]. In the travelling waist collision scheme, the optical focal point depends on the longitudinal position of a particle within the bunch. In other words, particles with different longitudinal coordinates in collision "see" the same and minimal beta functions. In circular colliders the travelling waist can be realized by a combination of accelerator components that provides a transformation described by a Hamiltonian $H=H_0-(zp_y^2)/2$ relating the longitudinal position $z$ and the vertical momentum $p_y$. For example, as proposed in [46], the travelling waist with the crab crossing can be obtained by using together crab cavities and sextupole magnets.

The longitudinal strong RF focusing is an alternative way to obtain short bunches at the IP [47]. It consists in realizing a large momentum compaction of the lattice together with a strong RF gradient. In this regime the bunch length is no longer constant, but it is modulated along the ring and can be minimized at the IP. In turn, if the main impedance generating elements of the ring are located where the bunch remains long, it is possible to minimize the strength of wake fields. This helps to avoid microwave instabilities and excessive bunch lengthening due to the potential well distortion. This concept was proposed as one of the possible options for the DAΦNE upgrade [54].

Contrary to the conventional strategy, the crab waist collision scheme requires small emittance, large Piwinski angle and larger crossing angle; there is no need to decrease the bunch length and push beam currents beyond the values already achieved in the present factories. At present this scheme is considered to be most attractive for the next generation lepton factories since it holds the promise of increasing the luminosity of the storage-ring colliders by 1-2 orders of magnitude beyond the current state-of-art. Let us discuss the crab waist collision concept in detail.

**Crab Waist Collision Scheme**

The CW scheme can substantially increase collider luminosity since it combines several potentially advantageous ideas. Let us consider two bunches colliding under a horizontal crossing angle $\theta$ (as shown in Fig. 1a). Then, the CW principle can be explained in the three basic steps. The **first one** is large Piwinski angle $\Phi = (\sigma_z/\sigma_x)tg(\theta/2) >> 1$. In the CW scheme the Piwinski angle is increased by decreasing the horizontal beam size and increasing the crossing angle. In this way we can gain in luminosity and the horizontal tune shift decreases (as described in previous Section); parasitic collisions (PC) become negligible since with higher crossing angle and smaller horizontal beam size the beam separation at the PC is larger in terms of $\sigma_x$. But the most important effect is that the overlap area of the colliding bunches is reduced, since it is proportional to $\sigma_x/\theta$ (see Fig. 2).

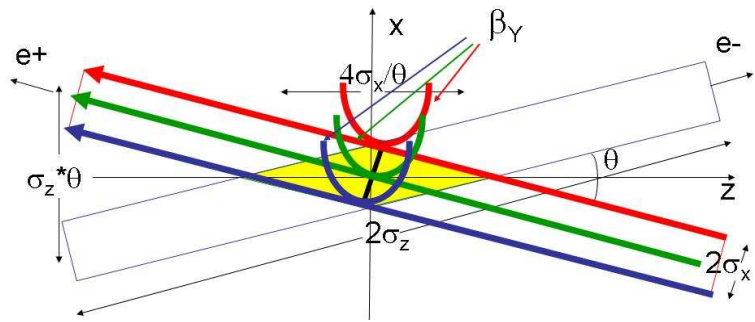

a) Crab sextupoles OFF.

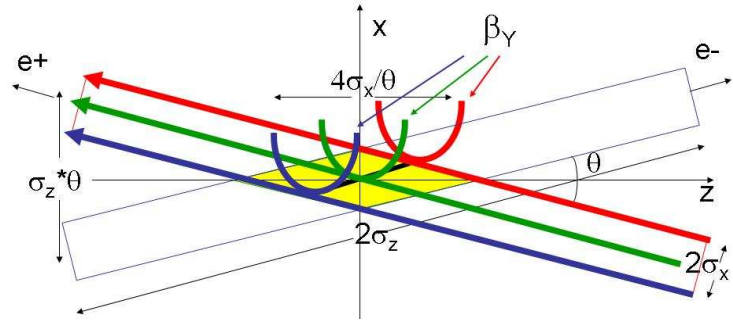

b) Crab sextupoles ON.

Figure 2: Crab Waist collision scheme.

Then, as the **second step**, the vertical beta function $\beta_y$ can be made comparable to the overlap area size (i.e. much smaller than the bunch length):

$$\beta_y^* \approx \frac{2\sigma_x}{\theta} \cong \frac{\sigma_z}{\Phi} << \sigma_z$$

So, reducing $\beta_y^*$ at the IP gives us several advantages:

- Luminosity increase with the same bunch current;
- Possibility of the bunch current increase (if it is limited by $\xi_y$), thus farther increasing the luminosity;
- Suppression of the vertical synchrobetatron resonances [55].

Besides, there are additional advantages in such a collision scheme: there is no need in decreasing the bunch length to increase the luminosity as required in standard collision schemes. This will certainly helps solving the problems of HOM heating, coherent synchrotron radiation of short bunches, excessive power consumption, etc.

However, implementation of these two steps introduces new beam-beam resonances which may strongly limit the maximum achievable tune shifts. At this point the crab waist transformation [31, 45] enters the game boosting the luminosity. This is the **third step**. As it is seen in Fig. 2b, the beta function waist of one beam is oriented along the central trajectory of the other one. In practice the CW vertical beta function rotation is provided by sextupole magnets placed on both sides of the IP in phase with the IP in the horizontal plane and at $\pi/2$ in the vertical one (as shown in Fig. 3).

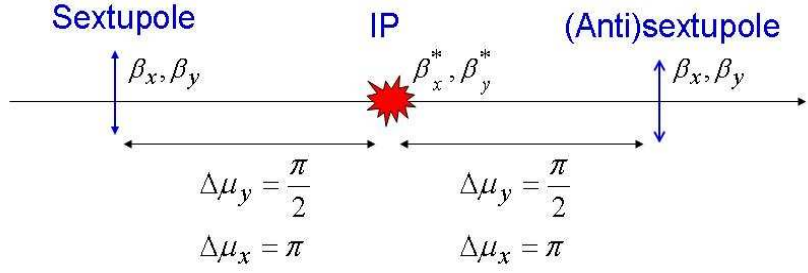

Figure 3. Crab sextupole locations.

The crab sextupole strength should satisfy the following condition depending on the crossing angle and the beta functions at the IP and the sextupole locations:

$$K = \frac{1}{\theta}\frac{1}{\beta_y^* \beta_y}\sqrt{\frac{\beta_x^*}{\beta_x}}$$

The crab waist transformation gives a small geometric luminosity gain due to the vertical beta function redistribution along the overlap area. It is estimated to be of the order of several percent. However, the dominating effect comes from the suppression of betatron (and synchrobetatron) resonances arising (in collisions without CW) due to the vertical motion modulation by the horizontal betatron oscillations [31, 44, and 45].

In order to understand the origin of the effect let us consider a simplified picture (Fig.4): a particle of one beam colliding with a thin opposite beam. Performing horizontal betatron oscillations the particle passes the crab sextupole at different horizontal offsets and is focused by the sextupole in such a way that:

- in collision it "sees" the same density of the opposite beam and has the same (minimum) vertical beta function. In other words, the strength of the vertical beam-beam kick does not depend on the horizontal coordinate;
- besides, it can be shown [31] that the vertical phase advance between the sextupole and the collision point always remains the same ($\Delta\mu_y = \pi/2$).

So, as we see, in crab waist collision the vertical motion is no longer affected by the horizontal oscillations.

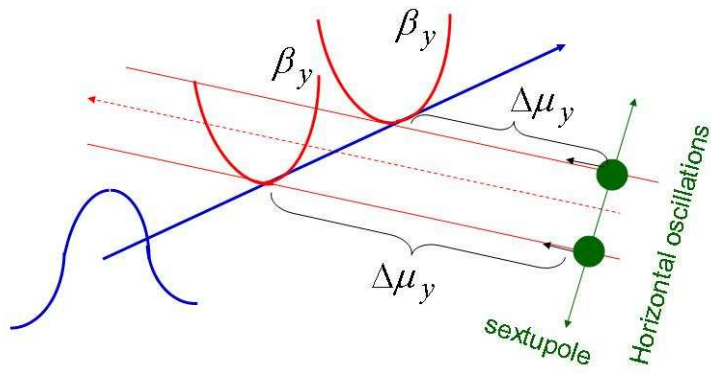

Figure 4. Particle's interaction with crabbed beam.

The effect of the resonance suppression can be easily demonstrated by performing numerical simulations of the beam-beam interaction. Figure 5 shows luminosity tune scans, i.e. the luminosity as a function of the horizontal ($v_x$) and vertical ($v_y$) normalized betatron frequencies, for the two typical cases: a) standard scheme of collisions with the low Piwinski angle $\Phi < 1$ and vertical beta function $\beta_y$ comparable with the bunch length $\sigma_z$ (as in KEKB and DAΦNE before upgrade) and b) crab waist collisions with large Piwinski angle $\Phi >> 1$ and $\beta_y$ comparable to the small overlap area $\sigma_x/\theta$ (as in SuperB, SuperC-Tau).

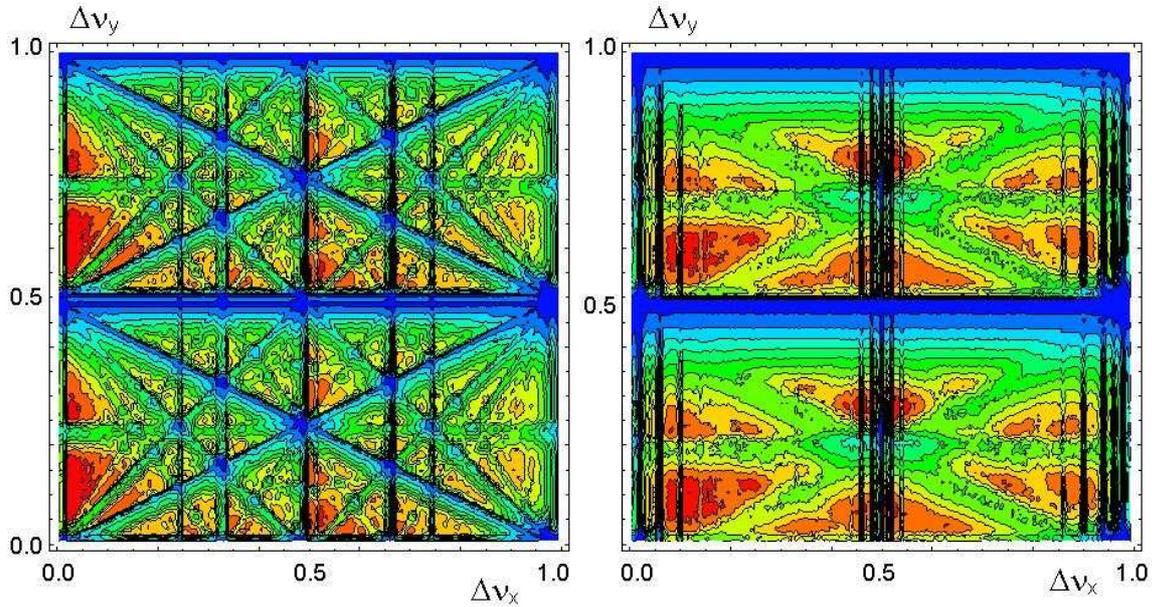

Figure 5. Luminosity scans for:
conventional collision scheme (left); crab waist collision scheme (right).

As it is seen in Fig.5, in crab waist collision X-Y coupling beam-beam resonances are successfully damped. Namely these resonances are considered to be most dangerous for flat colliding beams of the electron-positron factories. The beam-beam resonances can drive particle to higher oscillation amplitudes through different nonlinear mechanisms (see the reviews [56, 57] for details) thus leading to both beam core blow up and non-gaussian tail growth. Figure 6 shows a beneficial effect of the resonance suppression with crab waist sextupoles using an example of beam-beam simulations for the SuperB factory.

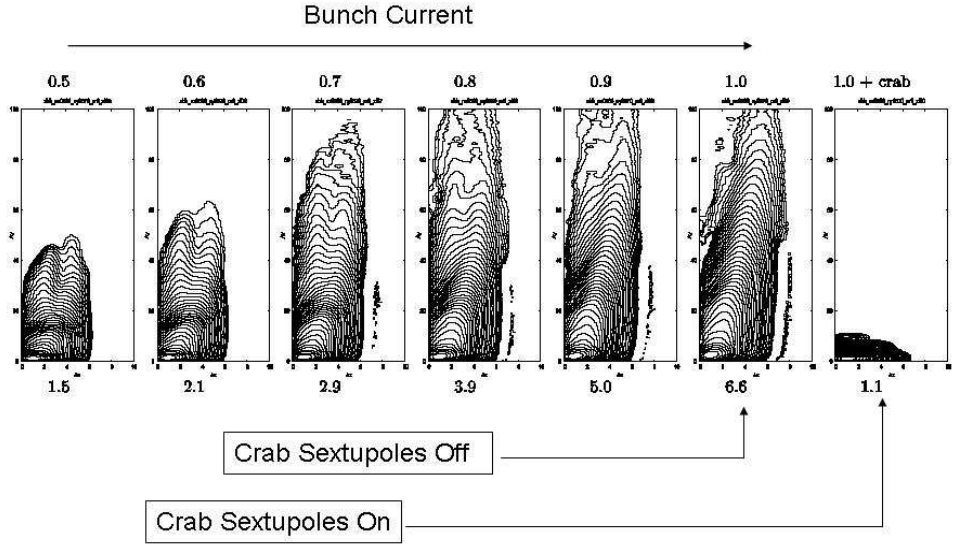

Figure 6. Charge density contour plots in the space of normalized betatron amplitudes for different normalized bunch currents (see the text for explanations).

Here we see a series of equilibrium charge density contour plots in the space of normalized betatron amplitudes, $A_x/\sigma_x$ and $A_y/\sigma_y$. The successive contour levels are at a constant ratio $e^{1/2}$ below each other. The numbers above each plot correspond to the normalized bunch current, where 1 means the nominal design current. In turn, the numbers below stand for the vertical beam size blow up $\sigma_y/\sigma_{y0}$, where $\sigma_{y0}$ is design vertical beams size without blow up. As we see, increasing the bunch current from 0.5 to 1 with the crab sextupoles switched off leads to the beam size blow up (a factor 6.6) and dramatic non-gaussian distribution tail growth. Instead, switching on the sextupoles (the last plot) practically eliminates the beam-beam blow up and suppresses the tails. In other worlds, one can expect a strong luminosity increase and beam lifetime improvement after switching on the crab sextupoles.

The crab waist collision scheme has been successfully tested at the electron-positron Φ-factory DAΦNE, providing luminosity increase by a factor of 3 [32], in a good agreement with numerical simulations.

**Crab Waist Test at DAΦNE**

DAΦNE is an electron-positron collider working at the c.m. energy of the Φ resonance (1.02 GeV c.m.) to produce a high rate of K mesons [25]. In its original configuration the collider consisted of two independent rings having two common Interaction Regions (IR) and an injection system composed of a full energy linear accelerator, a damping/accumulator ring and transfer lines. Figure 7 shows a view of the DAΦNE accelerator complex.

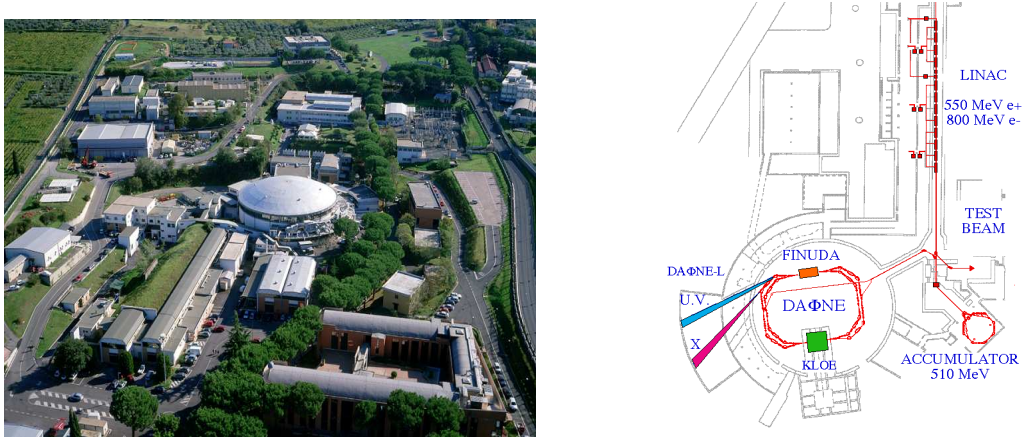

Figure 7. DAΦNE accelerator complex.

Since year 2000 till the middle 2007 DAΦNE was delivering luminosity to three experiments, KLOE, FINUDA and DEAR, steadily improving performances in terms of luminosity lifetime and backgrounds [58]. In these years the collider has undergone several progressive upgrades implemented during the shutdowns for detector changeover. The best machine performances were obtained in the KLOE and FINUDA runs. In particular, we reached a peak luminosity of $1.5\text{-}1.6 \times 10^{32}$ cm$^{-2}$s$^{-1}$ with a maximum daily integrated luminosity of about 10 pb$^{-1}$.

Table 3. DAΦNE main parameters (KLOE configuration)

| Energy | GeV | 0.51 |
|---|---|---|
| Circumference | m | 97.69 |
| RF frequency | MHz | 368.26 |
| Harmonic number | | 120 |
| Damping time (x, y) | turns | 110000 |
| Bunch length | cm | 1-3 |
| Emittance | mm x mrad | 0.34 |
| Coupling | % | 0.2-0.3 |
| Beta functions at IP (x, y) | m | 1.5/0.018 |
| Maximum tune shifts | | 0.03 |
| Number of bunches | | 111 |
| Maximum beam currents | A | 2.45/1.40 |

Table 3 shows the main DAΦNE parameters during the KLOE experimental run. As it is seen, despite the low energy, the long damping time in terms of revolution turns and a very short separation between consequtive bunches (2.7 ns), it was possible to accumulate very high intensity beams in collision, with 2.45 A in the electron beam and 1.40 A in the positron one.

In 2007, during a five month shut down used for the installation of the experimental detector SIDDHARTA, DAΦNE was upgraded implementing the crab waist collision scheme [59]. Table 4 shows a comparison of the main beam parameters at the interaction point (IP) for the DAΦNE upgrade with those of the previous runs for the KLOE and FINUDA experiments. As one can see from Table 4 the Piwinski angle was increased (by a factor of 4-5) and the collision region length reduced by doubling the crossing angle, decreasing the horizontal beta function almost by an order of magnitude and slightly decreasing the horizontal emittance. In turn, the vertical beta function at the IP was decreased by a factor of 2. The crab waist transformation was provided by two electromagnetic sextupoles installed at both ends of the experimental interaction region with the required phase advances between them and the IP. Their integrated gradient is about a factor 5 higher than that of normal sextupoles used for chromaticity correction.

Table 4. DAΦNE IP parameters

| Parameters | KLOE | FINUDA | SIDDHARTA |
|---|---|---|---|
| Date | September 2005 | April 2007 | June 2009 |
| $\varepsilon_x$, mm mrad | 0.34 | 0.34 | 0.25 |
| $\beta_x$, m | 1.5 | 2.0 | 0.25 |
| $\sigma_x$, mm | 0.71 | 0.82 | 0.25 |
| $\theta$, mrad | 25 | 25 | 50 |
| $\sigma_z$, cm | 2.5 | 2.2 | 1.7 |
| $\Phi$ | 0.44 | 0.34 | 1.70 |
| $\beta_y$, cm | 1.8 | 1.9 | 0.93 |

Right from the start of commissioning, the effectiveness of the new collision scheme was confirmed by several measurements and qualitative observations of the beam-beam behavior. The simplest and most obvious test consisted in switching off the crab waist sextupoles of one of the colliding beams. This blew up both horizontal and vertical transverse beam sizes of that beam and created non-gaussian tails of the beam distribution, seen on the synchrotron light monitors (see Fig. 8). At the same time, a luminosity reduction was recorded by all the luminosity monitors. This behavior is compatible with the prediction of additional beam-beam resonances when the crab sextupoles are off.

The best peak luminosity of $4.53 \times 10^{32}$ cm$^{-2}$s$^{-1}$ was obtained in June 2009 (see Fig. 9) together with a daily integrated luminosity exceeding 15pb$^{-1}$. As one can see from Fig. 9, the best present luminosity is by a factor 3 higher than that in the runs before the upgrade. The maximum peak luminosity is already very close to the design value of $5 \times 10^{32}$ cm$^{-2}$s$^{-1}$, and work is still in progress to achieve this ultimate goal. The vertical tune shift parameter has been significantly improved and it is now as high as 0.044 (a factor 1.5 higher than before). It is worth mentioning that in weak-strong collisions when the electron beam current is much higher than the positron one the tune shift has reached almost 0.09.

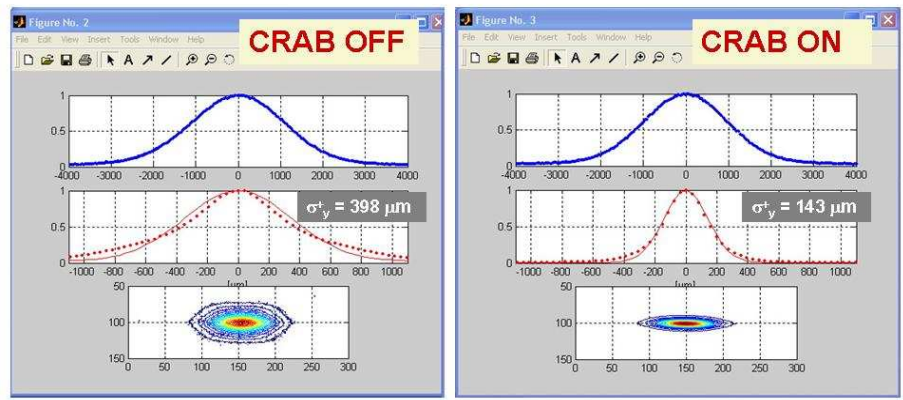

Figure 8: Transverse beam profiles with crab on and off.

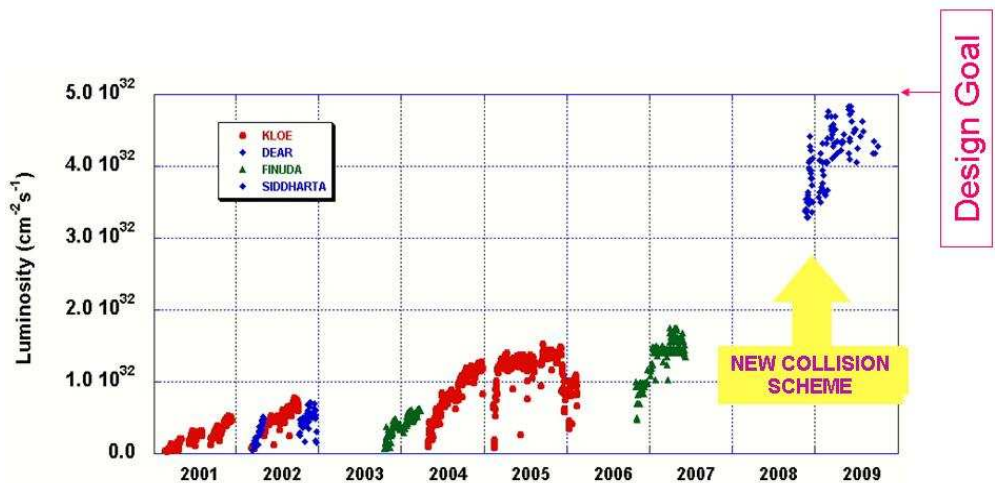

Figure 9: DAΦNE peak luminosity history.

A comprehensive numerical simulation study has been undertaken for comparison with the experimental data and test once more the effectiveness of the crab waist collision scheme [60, 61]. In turn, several dedicated experiments have been carried out at DAΦNE for the numerical codes benchmarking. In particular, we have found that the measured luminosity is only 15-20 % lower than predicted by the strong-strong self-consistent simulations with BBSS [62] and SBBE [63] codes. In our opinion, this is a good agreement given that the ideal strong-strong simulations do not take into account many factors, both single- and multibunch, affecting the luminosity such as: lattice nonlinearities, e-cloud effects, trapped ions, wake fields, gap transients, hardware noise etc.

A couple of experimental DAΦNE runs were dedicated to tune and to optimize the collider in the weak-strong regime in order to compare measured data with results of the weak-strong code LIFETRAC [64] modified to be able to simulate the crabbed strong beam. In order to eliminate the crosstalk between e-cloud effects and beam-beam interaction the stored positron beam was chosen to be the weak one (100-200 mA). All the parameters necessary for numerical simulations such as beam currents, transverse beam sizes, bunch length etc. were measured and recorded during these runs. As shown in Fig. 10, practically there is no difference between the numerical predictions and measured luminosity.

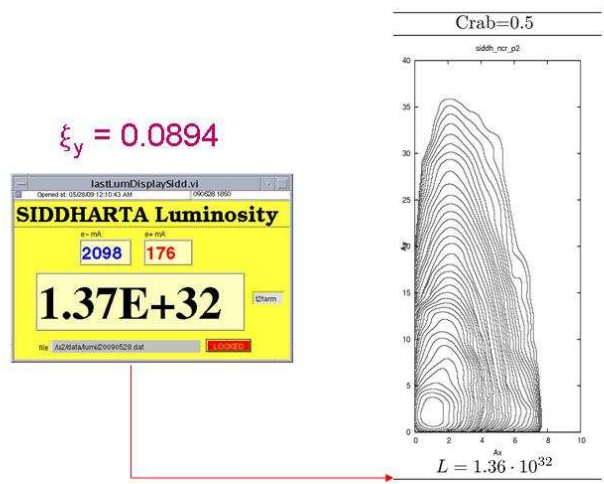

Figure 10: Comparison of measured (left picture - luminosity monitor display) and calculated (right picture) luminosity.

To complete the CW scheme studies with a kind of control experiment, several hours have been devoted to tuning the collider with the crab sextupoles off. Figure 11 shows a comparison of the luminosity as a function of beam current product obtained with the crab sextupoles on and off. The maximum luminosity reached in the latter case was only 1.6-1.7x10$^{32}$ cm$^{-2}$s$^{-1}$. It is worth remarking that another drawback becomes very important in collision without the crab sextupoles: besides much bigger vertical blow up leading to luminosity decrease, a sharp lifetime reduction was observed at single bunch currents as low as 8-10 mA. For this reason the red curve in Fig. 11 stops at much lower beam currents. Such a behavior is also consistent with numerical predictions based on beam-beam simulations taking into account realistic lattice nonlinearities [65].

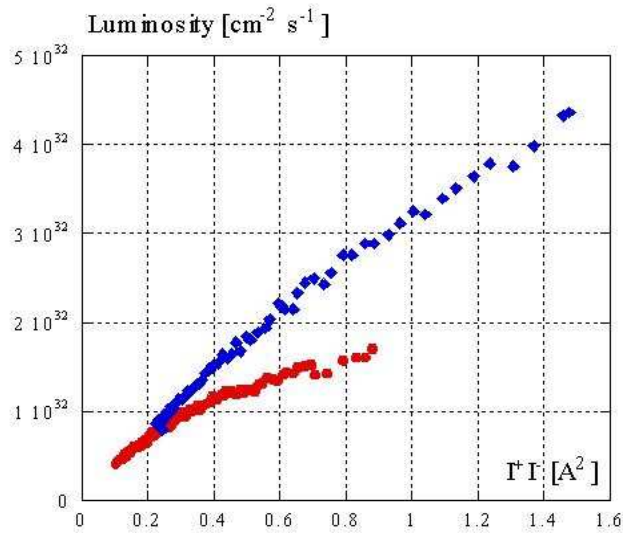

Figure 11. Measured luminosity as a function of beam current product
for crab sextupoles on (blue) and off (red).

**Future Electron-Positron Factories**

The successful test of crab waist collisions at DAΦNE and advantages of the crab waist collision scheme have triggered several collider projects exploiting its potential. In particular, physics and accelerator communities are discussing and developing new projects of a SuperB-factory [33, 34] and a SuperC-Tau factory [35] with luminosities about two orders of magnitude beyond those achieved at the present B- and Tau-Charm factories.

*1. SuperB*

The crab waist collision is the basic concept of the SuperB project [33, 66] aimed at the constructing of a very high luminosity asymmetric e+e- flavor factory with a possible location either near the campus of the University of Rome at Tor Vergata or at the site of the INFN Frascati National Laboratories. Figure 12 shows the SuperB layout at the Frascati (INFN LNF) site.

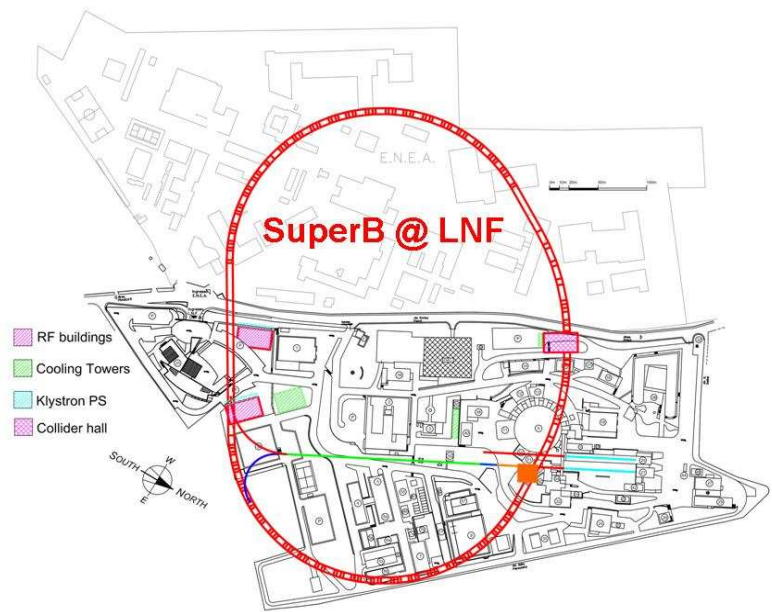

Figure 12. SuperB footprint at LNF.

The SuperB accelerator is being designed to satisfy the following requirements:

- Very high luminosity, $> 10^{36}$ cm$^{-2}$s$^{-1}$;
- Longitudinally polarized beam (e-) at IP (>80%);
- Ability to collide at charm threshold (3.8 GeV c.m.);
- Flexible parameter choice;
- Flexible lattice.

Column 1 of Table 5 shows the baseline parameter set that relies on the following criteria:

- to maintain wall plug power, beam currents, bunch lengths, and RF requirements comparable to present B-Factories, with parameters as close as possible to those achieved or under study for the ILC Damping Ring and at the ATF ILC-DR test facility;
- to reuse as much as possible of the PEP-II hardware;
- to simplify the IR design as much as possible, reducing the synchrotron radiation in the IR, HOM power and increasing the beam stay-clear;
- to eliminate the effects of the parasitic beam crossing, at the same time relaxing as much as possible the requirements on the beam demagnification at the IP;
- to design a Final Focus (FF) system to follow as closely as possible existing systems, and integrating it as much as possible into the ring design.

Table 5: SuperB parameters for baseline, low emittance and high current options, and for τ/charm running.

| Parameter | Units | Base Line | | Low Emittance | | High Current | | τ-charm | |
|---|---|---|---|---|---|---|---|---|---|
| | | HER (e+) | LER (e-) | HER (e+) | LER (e-) | HER (e+) | LER (e-) | HER (e+) | LER (e-) |
| LUMINOSITY | cm$^{-2}$ s$^{-1}$ | 1.00E+36 | | 1.00E+36 | | 1.00E+36 | | 1.00E+35 | |
| Energy | GeV | 6.7 | 4.18 | 6.7 | 4.18 | 6.7 | 4.18 | 2.58 | 1.61 |
| Circumference | m | 1258.4 | | 1258.4 | | 1258.4 | | 1258.4 | |
| X-Angle (full) | mrad | 66 | | 66 | | 66 | | 66 | |
| $\beta_x$ @ IP | cm | 2.6 | 3.2 | 2.6 | 3.2 | 5.06 | 6.22 | 6.76 | 8.32 |
| $\beta_y$ @ IP | cm | 0.0253 | 0.0205 | 0.0179 | 0.0145 | 0.0292 | 0.0237 | 0.0658 | 0.0533 |
| Coupling (full current) | % | 0.25 | 0.25 | 0.25 | 0.25 | 0.5 | 0.5 | 0.25 | 0.25 |
| Emittance x (with IBS) | nm | 2.00 | 2.46 | 1.00 | 1.23 | 2.00 | 2.46 | 5.20 | 6.4 |
| Emittance y | pm | 5 | 6.15 | 2.5 | 3.075 | 10 | 12.3 | 13 | 16 |
| Bunch length (full current) | mm | 5 | 5 | 5 | 5 | 4.4 | 4.4 | 5 | 5 |
| Beam current | mA | 1892 | 2447 | 1460 | 1888 | 3094 | 4000 | 1365 | 1766 |
| RF frequency | MHz | 476. | | 476. | | 476. | | 476. | |
| Number of bunches | # | 978 | | 978 | | 1956 | | 1956 | |
| Tune shift x | | 0.0021 | 0.0033 | 0.0017 | 0.0025 | 0.0044 | 0.0067 | 0.0052 | 0.0080 |
| Tune shift y | | 0.097 | 0.097 | 0.0891 | 0.0892 | 0.0684 | 0.0687 | 0.0909 | 0.0910 |
| Total RF Wall Plug Power | MW | 16.38 | | 12.37 | | 28.83 | | 2.81 | |

The machine is designed to have flexibility for the parameters choice with respect to the baseline: the horizontal emittance can be decreased by a factor of ~2 in both rings by changing the partition number (by changing the RF frequency, as done in LEP, or the orbit in the arcs) and the natural emittance by readjusting β functions.

Moreover the FF system has a built-in capability for decreasing the IP β functions of a factor of ~2, and the RF system will be able to support higher beam currents than the baseline, when all the available PEP RF units will be installed.

Based on these considerations, columns 2 and 3 in Table 5 show different parameters options:

- "Low Emittance" case relaxes RF requirements and problems related to high current operations (including wall-plug power) but puts more strain on the optics and the tuning capabilities;

- "High Current" case relaxes requirements on vertical emittance and IP β functions, but high currents issues are enhanced in terms of instabilities, HOM, synchrotron radiation, wall-plug power, etc.

The cases considered have several parameters kept as much constant as possible (bunch length, IP stay clear etc…), in order to reduce their impact on other unwanted effects (Detector background, HOM heating etc…).

SuperB can also operate at lower cm energy (|/charm threshold energies near 3.8 GeV) with a somewhat reduced luminosity and minimal modifications to the machine: the beam energies will be scaled, maintaining the nominal energy asymmetry ratio used for operation at the cm energy of the ∏(4S). The last column in Table 5 shows preliminary parameters for the run at the |/charm.

*Rings Lattice*

The SuperB HER and LER ring lattices need to comply with several constraints. First of all extremely low emittances and IP beam sizes, needed for the high luminosity, damping times, beam lifetimes and polarization for the electron beam. The rings can be basically considered as two Damping Rings (similar to ILC and CLIC ones) with the constraint to include a FF section for collisions. So, the challenge is not only how to achieve low emittance beams but how to choose the other beam parameters to be able to reach design luminosity with reasonable lifetimes and small beams degradation. For this purpose a new "Arc cell" design has been adopted for SuperB [67]. The extremely low-β in the FF system, together with the Crab Waist scheme, requires a special optics that provides the necessary beam demagnification at the IP, corrects its relative chromaticity and provides the necessary conditions and constraints for the "Crab Waist" optics.

Both rings are located in the horizontal plane. The FF is combined with the two arcs in two half-rings (one inner, one outer) and a straight section on the opposite side, which comes naturally to close the ring and readily accommodate the RF system and other necessities (e.g. injection). In this utility region crossing without collisions for the two rings will be provided. More details on the lattice can be found in Ref [67].

*Interaction Region*

The high luminosity is achieved primarily with the implementation of very small $\beta_x^*$ and $\beta_y^*$

values at IP. These conditions are principal driving terms in the design of the IR. The FF doublet (QD0 and QF1) must be as close as possible to the IP in order to minimize chromatic and other higher-order aberrations from these magnet fields. The present IR design with a crossing angle of +/-33 mrad uses separate focusing elements for each beam. The QD0 magnet is now a twin design of side-by-side super-conducting quadrupoles. The magnet windings are designed so that the fringe field of the neighbouring magnet can be cancelled maintaining high quality quadrupole fields for both beams. Further details about the IR design can be found in the Ref [68].

*Polarization*

SuperB will achieve polarized beams by injecting polarized electrons into the LER. We chose the LER rather than the HER because the spin rotators employ solenoids which scale in strength with energy.

In SuperB at high luminosity the beam lifetime will be only 3…5 minutes and continuous-injection ("trickle-charge") operation is a key component of the proposal. By injecting at a high rate with a polarized beam one can overcome the depolarization in the ring as long as the spin diffusion is not too rapid. In the ring arcs the polarization must be close to vertical to minimize depolarization. In order to obtain longitudinal polarization at the IP, a rotation of the spin by 90° about the radial axis is required. A rotation of 90° in a solenoid followed by a spin rotation of 90° in the horizontal plane by dipoles also provides the required net rotation about the radial axis without vertical bending and was therefore adopted. The solenoid field integral required is 21.88 Tm for 90° spin rotation, well within the technical capabilities of superconducting solenoids of the required aperture. After the IP, the polarization has to be restored to vertical by a second spin rotator. Due to the low beam lifetime, it turns out that a symmetric spin-rotator scheme is feasible and can achieve 70% polarization or better. More details on these studies can be found in Ref [69].

*Injection System*

The injection system for SuperB [70] is capable of injecting electrons and positrons into their respective rings at full energies. The HER requires positrons at 6.7 GeV and the LER 4.18 GeV polarized electrons. At full luminosity and beam currents, up to 4 A, the HER and LER have expected beam lifetimes in the range 3÷5 minutes. Thus, the injection process must be continuous, to keep nearly constant beam current and luminosity. Multiple bunches are injected on each linac pulse into one or the other of the two rings. Electrons from the gun source are longitudinally polarized: the spins are rotated to the vertical plane in a special transport section downstream of the gun. The spins then remain vertical for the rest of the injection system and injected in this vertical state into the LER. Positron bunches are generated by striking a high charge electron bunch onto a positron converter target and collecting the emergent positrons. Electron to positron conversion is done at about 0.6 GeV using a newly designed capture section to produce a yield of more than 10% [71]. The transverse and longitudinal emittances of both beams are larger than the LER and HER acceptances and must be pre-damped. A specially designed Damping Ring at 1 GeV, shared by both beams to reduce costs, is used to reduce the injected beam emittances.

*2. SuperKEKB*

SuperKEKB is another Super B-factory project [34]. It is a natural upgrade of the very successful KEKB, Japanese B-factory at KEK (Tsukuba) [27]. The design luminosity goal of the project is $0.8 \times 10^{36}$ cm$^{-2}$s$^{-1}$, i.e. by a factor 40 higher than the world record luminosity of $2.1 \times 10^{34}$ cm$^{-2}$s$^{-1}$ achieved at KEKB. Figure 13 shows the SuperKEKB layout.

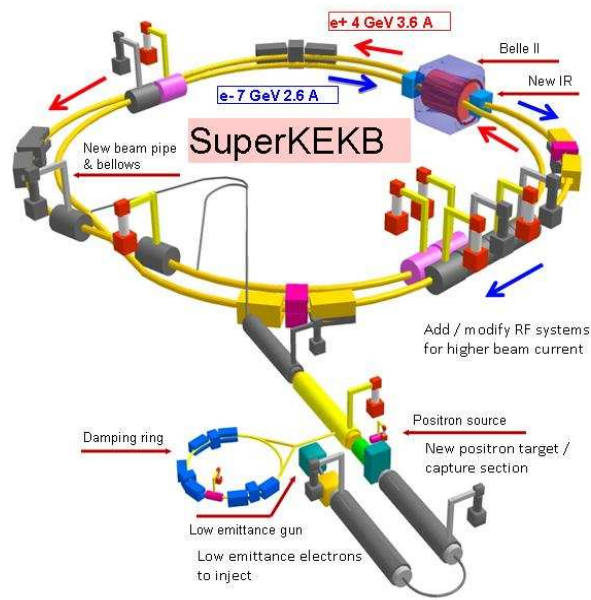

Figure 13. SuperKEKB schematic view.

Initially the upgrade was planned to follow the standard "brute-force" approach (so- called "High Current Option") based on:

- beam current increase by a factor of 3 to 5 with respect to the already achieved values;
- a very high beam-beam tune shift parameter, as high as 0.30;
- lower beta functions at the interaction point and, respectively shorter bunch length;

- crab crossing collisions (exploiting crab cavities).

However, it was recognized that such a scheme has several drawbacks and issues to study and solve such as:

- the assumed short bunches with $\sigma_z = 3$ mm would emit the coherent synchrotron radiation (CSR) that can results in excessive power loss, beam quality degradation and eventual beam instabilities;
- the vertical beam-beam tune shift parameter presently achieved in crab crossing collision at KEKB is about 0.09, which is far smaller that expected numbers, 0.15 for KEKB and 0.30 for SuperKEKB;
- the huge design beam currents require dedicated R&D studies for various vacuum chamber components and investigation of measured for different instability cures;
- high construction and operation costs.

So, it has been decided to abandon the High Current approach and to follow Italian strategy (CW) [72]. Now, as it can be seen in Table 6, beam parameters at the interaction point are very similar to those of the SuperB baseline design (see Table 5 for comparison). Due to the small beam sizes at IP the current upgrade option is called "Nano-Beam Scheme".

Table 6. SuperKEKB parameters

| Parameters | Units | SuperKEKB | |
| --- | --- | --- | --- |
| | | HER (e-) | LER (e+) |
| Circumference, $C$ | m | 3016.3 | 3016.3 |
| Energy, $E$ | GeV | 7 | 4 |
| Crossing angle, $\theta$ | mrad | 83 | |
| $\beta_x$ at IP | cm | 2.4 | 3.2 |
| $\beta_y$ at IP | cm | 0.041 | 0.027 |
| Emittance, $\varepsilon_x$ | nm | 2.4 | 3.1 |
| Coupling | % | 0.35 | 0.40 |
| Bunch length, $\sigma_z$ | mm | 5 | 6 |
| Beam current, $I$ | mA | 2620 | 3600 |
| $\sigma_x$ at IP | μm | 7.75 | 10.2 |
| $\sigma_y$ at IP | μm | 0.059 | 0.059 |
| Hor. tune shift, $\xi_x$ | | 0.0028 | 0.0028 |
| Vert. tune shift, $\xi_y$ | | 0.0875 | 0.0900 |
| Luminosity | cm$^{-2}$s$^{-1}$ | $0.8 \times 10^{36}$ | |

The design luminosity of $0.8 \times 10^{36}$ cm$^{-2}$s$^{-1}$ is to be achieved by exploiting the first two steps of the CW concept: with respect to KEKB the large Piwinski angle is obtained by drastic horizontal emittance and horizontal beta function reduction and the crossing angle increase till 66 mrad, while the vertical beta function at the IP is squeezed down to 027 (0.41) mm in LER (HER), proportionally to the collision area length reduction. Besides, it is thought that crab waist sextupoles may bring another bonus on the luminosity by a factor of more than 2. The work on dynamic aperture optimization with the CW sextupoles is in progress. It is worthwhile noting also that a conservative value of the beam-beam parameter of 0.09 is assumed for the project.

With respect to Italian SuperB the SuperKEKB project will not use beam polarization and it does not foresee the possibility to decrease the collider energy down to the charm threshold. However, SuperKEKB has important advantages: it will reuse the existing KEKB hardware and infrastructures (Linac, tunnels, buildings, technical services etc.) and it is already partially funded. Nevertheless, there is a long list of items to be upgraded or newly designed:

- New positron damping ring and new positron target [73]
- New RF gun for electrons with reduced emittance [74]
- New antechamber beam pipes for both rings [75]
- Al (Cu) beam pipes for LER (HER) [75]
- Mitigation techniques for e-cloud suppression [76, 77]
- New interaction region optics [78]
- New superconducting/permanent magnets around IP [78]
- etc.

*3. SuperC-Tau factory*

Budker Institute of Nuclear Physics (Novosibirsk, Russia) is promoting the project of a new generation SuperC-Tau factory (SCT) [35, 79, and 80]. The crab waist concept should allow reaching the project luminosity of $1\text{-}2 \times 10^{35}$ cm$^{-2}$s$^{-1}$ that is by more than 2 orders of magnitude higher than the luminosity $3.3 \times 10^{32}$ cm$^{-2}$s$^{-1}$ presently achieved at the operating τ-Charm factory BEPCII in Beijing [81].

The collider experimental program is aimed at the following studies [82]:

- D-Dbar mixing

- CP violation search in charm decays

- Study or rare and forbidden charm decays

- Standard Model tests in tau lepton decays

- Searching for lepton flavor violation

- CP/T violation search in tau lepton decays

- Production of polarized anti-nucleons

In order to fulfill these tasks, in addition to the high luminosity requirement, the SCT factory should provide collisions in the energy range between 2 and 4.5 GeV (c.m), foresee longitudinal polarization at the interaction point, be able to measure the energy with high precision etc. A schematic view of the SCT is shown in Fig. 14.

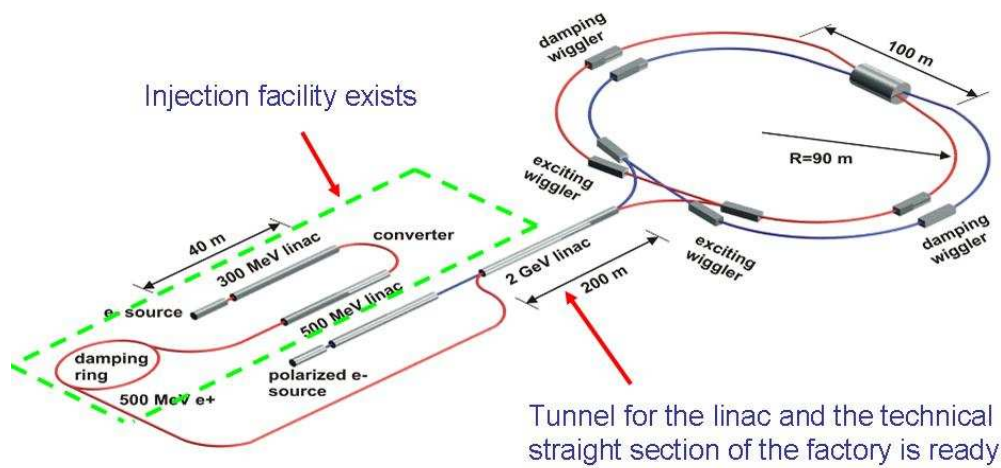

Figure 14. SuperC-Tau factory schematic layout.

The collider consists of a full energy injection system and two main storage rings having a single collision point. The rings have racetrack geometry with two arcs for beam bending and required emittance production; a straight interaction region (IR) with CW optics and sextupoles for local chromaticity correction; and a long straight section opposite to IR for beam injection, RF cavities and other technical equipment installation. Several straight insertions are foreseen for damping wigglers that are necessary to keep the high luminosity in the whole energy range. Polarization manipulation is provided by a system of Siberian Snakes.

The positron injection system will rely on the existing BINP injection facility [83]. After a moderate upgrade it will be able to provide required positron beam intensities in the top-up injection mode [80]. The electron beam injection chain starts with a Polarized Electron Source (PES) followed by a 510 MeV linac. The PES will be similar to the one developed by BINP and successfully operated for many years at AmPS (Netherlands) [84]. At the final stage a common 200 m long linac will accelerate both electrons and positron from 510 MeV to the nominal collider energy. It is worthwhile mentioning that the tunnel for the linac and the technical straight section of the main rings has already beam constructed.

The peak luminosity has been optimized for the beam energy of 2 GeV. Table 7 shows some SCT factory parameters for this energy and compares them with the design parameters of the Chinese τ-charm factory.

Table 7. SuperC-Tau and BEPCII design parameters

| Parameters | | BEPCII (design) | SuperC-Tau |
| --- | --- | --- | --- |
| Energy | $E$, GeV | 1.89 | 2.00 |
| Circumference | $C$, m | 238 | 767 |
| Damping time | $\tau_x/\tau_y/\tau_z$, ms | 25/25/12.5 | 30/30/30 |
| Beam current | $I$, A | 0.91 | 1.68 |
| Number of bunches | $n_b$ | 93 | 384 |
| Energy spread | $\sigma_E$ | 5.15x10$^{-4}$ | 7.10x10$^{-4}$ |
| Bunch length | $\sigma_z$, cm | 1.5 | 0.9 |
| Beta functions | $\beta_x^*/\beta_y^*$, m | 1/0.015 | 0.04/0.0008 |
| Emittances | $\varepsilon_x/\varepsilon_y$, nm-rad | 144/2.2 | 8/0.04 |
| Beam sizes at IP | $\sigma_x/\sigma_y$, μm | 380/5.7 | 17.9/0.179 |
| Crossing angle | $\theta$, mrad | 22 | 60 |
| Piwinski angle | $\Phi$ | 0.435 | 15.1 |
| Tune shifts | $\xi_x/\xi_y$ | 0.04/0.04 | 0.13/0.0044 |
| Luminosity | $L$, cm$^{-2}$s$^{-1}$ | 1.0x10$^{33}$ | 1.1x10$^{35}$ |

As it can be seen, following the crab waist strategy the Piwinski angle in SCF is chosen to be 15.1, i.e. by a factor 35 larger than in BEPCII. This is achieved by using a factor of 3 larger crossing angle and much smaller horizontal beam sizes at the interaction point with respect to the BEPCII design. Accordingly, also the vertical beta function is much smaller than $\beta_y$ in the Beijing collider.

It worth mentioning here that the SCF parameters are more conservative with respect to those of SuperB and SuperKEKB projects: smaller total beam currents, larger transverse emittances, longer bunches, bigger beta functions at the interaction point. The only challenging parameter is the vertical tune shift as high as 0.13. However, performed numerical simulations [80] have shown that there are wide working point areas where the design luminosity and the high tune shift can be achieved without beam blowup and beam lifetime degradation.

At present the work is in progress on final focus and lattice improvement, dynamic aperture optimization, beam dynamics studies, Touschek lifetime increase etc.

**Conclusions**

The present generation of electron-positron factories was very successful in accumulating the record beam currents, achieving very high luminosities and developing accelerator physics and technology. However, the particle physics have required pushing the luminosity of storage-ring colliders further to unprecedented levels since this opens up unique opportunities for precision measurements of rare decay modes and extremely small cross section, which are sensitive to new physics beyond the Standard Model.

Several novel collision concepts and new collision schemes have been proposed, and some of them tested experimentally, to provide such a qualitative step in luminosity increase. At present the crab waist collision scheme is considered to be most prominent for the next generation factories since it holds the promise of increasing the luminosity of the storage-ring colliders by 1-2 orders of magnitude beyond the current state-of-art, without any significant increase in beam current and without reducing the bunch length.

The successful test of crab waist collisions at DAΦNE, Italian Φ-factory, and advantages of the crab waist collision scheme have triggered several collider projects exploiting its potential. In particular, physics and accelerator communities are discussing new projects of a SuperB-factory in Italy, SuperKEKB-factory in Japan and a Super-Tau-Charm factory in Novosibirsk with luminosities about two orders of magnitude beyond those achieved at the present B- and Tau-Charm factories. The design studies of the new generation particle Factories are in a very advanced stage.